\documentclass[12pt]{article}

\usepackage{amsmath,amssymb}
\usepackage{graphicx}
\usepackage[colorlinks=true,linkcolor=blue,citecolor=blue,urlcolor=blue]{hyperref}
\usepackage{booktabs}
\usepackage{xcolor}
\usepackage{tikz}
\usetikzlibrary{arrows.meta, positioning, shapes.geometric, fit, calc}
\usepackage[numbers,sort&compress]{natbib}
\usepackage[margin=1in]{geometry}
\usepackage{url}

\title{Agentic AI and Occupational Displacement: A Multi-Regional Task Exposure Analysis of Emerging Labor Market Disruption}

\author{Ravish Gupta$^{1}$\thanks{Corresponding author: \texttt{ravishgupta@ieee.org}} \and Saket Kumar$^{2}$ \\[0.5em]
\small $^{1}$AI Lead, BigCommerce; IEEE Senior Member \\
\small $^{2}$University at Buffalo, The State University of New York, Buffalo, NY, USA}

\date{}

\begin{document}

\maketitle

\begin{abstract}
This paper extends the Acemoglu--Restrepo task exposure framework to address the labor market effects of agentic artificial intelligence systems: autonomous AI agents capable of completing entire occupational workflows rather than discrete tasks. Unlike prior automation technologies that substitute for individual subtasks, agentic AI systems execute end-to-end workflows involving multi-step reasoning, tool invocation, and autonomous decision-making, substantially expanding occupational displacement risk beyond what existing task-level analyses capture. We introduce the Agentic Task Exposure (ATE) score, a composite measure computed algorithmically from O*NET task data using calibrated adoption parameters---not a regression estimate---incorporating AI capability scores, workflow coverage factors, and logistic adoption velocity. Applying the ATE framework across five major United States technology regions (Seattle--Tacoma, San Francisco Bay Area, Austin, New York, and Boston) over a 2025--2030 horizon, we find that 93.2\% of the 236 analyzed occupations across six information-intensive SOC groups (financial, legal, healthcare, healthcare support, sales, and administrative/clerical) cross the moderate-risk threshold (ATE $\geq$ 0.35) in Tier~1 regions by 2030, with credit analysts, judges, and sustainability specialists reaching ATE scores of 0.43--0.47. We simultaneously identify seventeen emerging occupational categories benefiting from reinstatement effects, concentrated in human-AI collaboration, AI governance, and domain-specific AI operations roles. Our findings carry implications for workforce transition policy, regional economic planning, and the temporal dynamics of labor market adjustment. This work provides the first multi-regional empirical application of the task exposure framework specifically targeting agentic AI systems and contributes a reproducible scoring methodology for future labor market monitoring.
\end{abstract}

\noindent\textbf{Keywords:} Agentic AI; Task Exposure; Occupational Displacement; Labor Market; Workforce Transition; Human-AI Collaboration; O*NET; Multi-Agent Systems

\bigskip

\section{Introduction}
\label{sec:introduction}

Economists have been predicting automation-driven job losses for decades, and the predictions keep being directionally right but temporally wrong. Industrial robots did displace manufacturing workers, but not on the timeline or at the scale that the initial forecasts implied \cite{autor2015why}.  When generative large language models arrived after 2022, the displacement conversation shifted abruptly upmarket: suddenly it was lawyers, analysts, and radiologists in the crosshairs, occupations that the prior literature had treated as automation-proof \cite{eloundou2023gpts}.

Yet almost all of these forecasts, including those built on the now-standard Autor--Levy--Murnane task framework \cite{autor2003skill}, share a quiet assumption: automation picks off tasks one at a time, and the occupation survives as a coordination shell around the remaining human work.  A contract lawyer loses the drafting subtask to an LLM; the lawyering function persists.

Agentic AI demolishes that assumption. An agent does not nibble at one subtask and hand control back. It receives a goal (``review this lease for non-standard indemnification clauses and draft a redline''), selects tools, executes a multi-step plan, checks its own output, and delivers a finished work product \cite{wang2024survey}. No intermediate human touch. When that capability exists, the economic logic changes. If an agentic bookkeeping system can close the monthly books at 80\% of the quality of a human bookkeeper, and a single human reviewer can catch the remaining errors across fifty such agents, the organizational incentive is not to redeploy bookkeepers to the leftover 20\% of tasks. It is to eliminate the role and staff a small exception-handling team.

This paper makes three contributions. First, we introduce the Agentic Task Exposure (ATE) score, a composite metric that extends the Acemoglu--Restrepo task displacement framework \cite{acemoglu2019automation} to account for end-to-end workflow coverage, current agentic AI capability benchmarks, and enterprise adoption velocity. Second, we apply the ATE framework empirically to 236 Standard Occupational Classification (SOC) codes spanning six major occupational groups, generating model-derived displacement projections across five major United States technology regions for the 2025--2030 period. Third, we identify occupational categories where agentic AI creates reinstatement effects: new roles that did not exist before agentic deployment creates demand for human oversight, governance, and collaboration at scale.

The five regions under study are San Francisco Bay Area, Seattle--Tacoma--Bellevue, Austin--Round Rock, New York--Newark, and Boston--Cambridge. These metros rank among the leading US centers for AI activity, collectively representing a substantial share of national AI job postings, patent filings, and venture capital investment \cite{brookings2024ai,maslej2024ai} and represent the early-adoption frontier from which labor market effects will diffuse nationally. The San Francisco Bay Area, with the highest concentration of AI venture capital and patent activity, serves as the Tier~1 leading indicator: its adoption curve is approximately 6--12 months ahead of Tier~2 hubs (Seattle, Austin, Boston) and 18 months ahead of Tier~3 metros (New York). Seattle is of particular interest as a Tier~2 region anchored by Amazon Web Services, Microsoft, and a dense ecosystem of AI-native companies. Its workforce effects in 2026--2027 preview what the broader national non-technical workforce will experience in 2028--2030, following the pattern established by the San Francisco Bay Area in 2024--2025.

The remainder of this paper is organized as follows. Section~\ref{sec:background} reviews related work on task-based automation frameworks and prior AI labor market analyses. Section~\ref{sec:framework} formalizes the ATE score and its components. Section~\ref{sec:data} describes data sources and methodology. Section~\ref{sec:results} reports empirical results across occupations and regions. Section~\ref{sec:reinstatement} analyzes reinstatement effects and emerging roles. Section~\ref{sec:policy} discusses policy implications. Section~\ref{sec:conclusion} concludes.

\section{Background and Related Work}
\label{sec:background}

\subsection{Task-Based Frameworks for Automation Analysis}

Autor, Levy, and Murnane \cite{autor2003skill} partitioned occupational tasks along two axes: routine versus non-routine, and cognitive versus manual. They argued that automation substitutes for the routine while complementing the non-routine. That two-by-two grid organized a generation of labor economics. Acemoglu and Restrepo then cast the intuition in production-theoretic terms \cite{acemoglu2019automation,acemoglu2020robots}: output results from a continuum of tasks allocated between capital and labor, so that the labor share rises or falls depending on whether new-task creation (the reinstatement channel) outpaces task displacement. Their contribution was to make the net employment effect endogenous rather than assumed.

Frey and Osborne \cite{frey2017future} took the occupation itself as the unit of analysis, asking experts to score 702 US occupations on overall ``computerizability.'' The headline result (47\% of occupations at high risk) attracted enormous media attention and equally intense methodological pushback. Arntz, Gregory, and Zierahn \cite{arntz2016risk} showed that once the analysis drops to the task level and accounts for task heterogeneity \emph{within} nominally exposed occupations, high-risk estimates fall to 9--14\% across OECD countries. The debate is unresolved, but the field converged on a practical consensus: task-level decomposition is necessary; occupation-level labels are too coarse.

Two more recent indices are central to our own calibration. Felten, Raj, and Seamans \cite{felten2021occupational} built the AI Occupational Exposure (AIOE) score by linking published AI capability advances to the O*NET ability taxonomy. Eloundou et al.\ \cite{eloundou2023gpts} constructed a parallel measure focused on GPT-4, combining crowd-worker annotations with GPT-4 self-assessments at the task level. Their striking finding was that 80\% of US workers hold jobs where at least 10\% of tasks are LLM-exposed, with the heaviest exposure concentrated among high-wage, degree-requiring occupations. This upended the ``routine-biased technical change'' narrative that had dominated since the early 2000s.

\subsection{Limitations of Existing Frameworks for Agentic AI}

Every framework cited above shares an architectural blind spot: they model automation as picking off individual tasks while the occupation stays intact, a residual structure for human labor to fill. That assumption is reasonable for narrow AI (an image classifier, a machine-translation engine, a code-completion plugin), each of which does one thing and hands control back. Agentic systems violate the assumption wholesale. They chain tool calls, maintain persistent memory across steps, plan multi-action sequences, and self-correct when intermediate outputs fail quality checks \cite{wang2024survey}.

Consider a concrete example familiar to any legal operations manager. A task-exposure analysis might flag 80\% of a paralegal's tasks as LLM-exposed yet conclude the occupation is safe because coordinating those tasks demands human judgment: knowing which clause to research first, when to escalate to the partner, how to handle a client who changes the fact pattern mid-memo. An agentic system collapses that reasoning. It receives the research goal, retrieves case law, drafts the memorandum, cross-checks citations against Westlaw, revises to comply with the partner's style guide, and delivers without intermediate hand-offs. The coordination bottleneck that protected the paralegal role is now inside the agent's planning module.

What is missing, then, is a measure of whether the full \emph{workflow composition} of an occupation sits inside agentic AI's operational envelope.  We formalize this as the workflow coverage factor below.

\subsection{Agentic AI: Definitions and Capabilities}

We define an agentic AI system as one that satisfies four criteria: (1) it receives a goal specification rather than a task specification; (2) it autonomously selects and invokes tools to pursue the goal; (3) it maintains state across multiple inference steps; and (4) it produces a completed work product without human intervention at intermediate steps. This definition encompasses current enterprise deployments of systems built on frameworks such as the Model Context Protocol (MCP) \cite{anthropic2024mcp}, LangGraph, and AutoGen, as well as multi-agent architectures in which specialized agents collaborate on complex goals.

The capability frontier moved fast in 2023--2024. GPT-4-class models reached at or above passing-score thresholds on the bar exam and the US Medical Licensing Exam, and came within a few percentage points of passing CFA Level~I \cite{openai2023gpt4}. More relevant to our framework: agentic extensions of these models began completing multi-step software engineering tasks on SWE-bench, executing end-to-end financial analyses, and resolving customer service tickets without human escalation. By mid-2024 this was no longer a research demo. Major enterprise technology vendors (including Amazon, Microsoft, Salesforce, ServiceNow, and Workday) disclosed production agentic deployments touching white-collar workflows in their 2024 earnings calls and investor presentations \cite{wang2024survey,mckinsey2023generative}. These disclosures indicate the kind companies make only when deployments are already affecting headcount planning.

\section{Theoretical Framework: Agentic Task Exposure}
\label{sec:framework}

\subsection{The Acemoglu--Restrepo Foundation}

Let $\mathcal{N}$ denote the set of tasks required to produce value in an economy, with a subset $\mathcal{N}_L \subset \mathcal{N}$ performed by labor and $\mathcal{N}_K \subset \mathcal{N}$ performed by capital. Automation technologies expand $\mathcal{N}_K$ at the expense of $\mathcal{N}_L$, creating the displacement effect. New task creation expands $\mathcal{N}$ at the boundary where labor holds comparative advantage, creating reinstatement. Following Acemoglu and Restrepo \cite{acemoglu2019automation}, the labor share in value added is:
\begin{equation}
s_L = \frac{W \cdot L}{Y} = f\!\left(\frac{|\mathcal{N}_L|}{|\mathcal{N}|}\right)
\label{eq:laborshare}
\end{equation}
where $W$ is the wage rate, $L$ is employment, and $Y$ is output. Automation that expands $\mathcal{N}_K$ reduces $s_L$ through the displacement effect; new task creation that expands $\mathcal{N}_L$ restores $s_L$ through the reinstatement effect.

\subsection{The Agentic Task Exposure Score}

We extend this framework by introducing the Agentic Task Exposure (ATE) score, which measures the degree to which an occupation $o$ is exposed to end-to-end displacement by current agentic AI systems. ATE scores are computed algorithmically from O*NET task data using calibrated adoption parameters; they are not regression estimates. For occupation $o$ with task set $\mathcal{T}_o$ drawn from O*NET's Task Statements database, the ATE score is:
\begin{equation}
\mathrm{ATE}_o(r, \tau) = \sum_{t \in \mathcal{T}_o} w_{o,t} \cdot \mathrm{CAP}(t) \cdot \mathrm{COV}(t,o) \cdot V(t, r, \tau)
\label{eq:ate}
\end{equation}

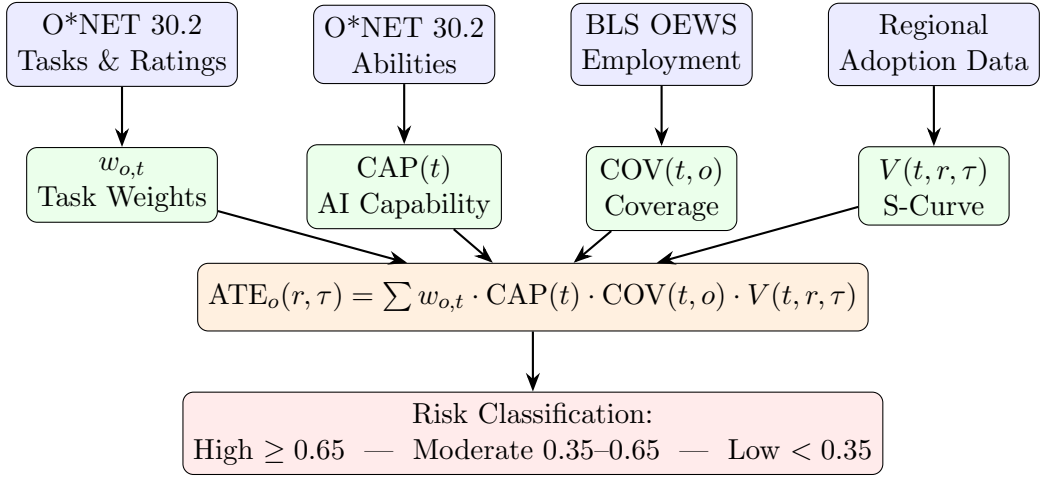
\begin{figure}[h!]
\centering
\begin{tikzpicture}[
    box/.style={draw, rounded corners, fill=blue!8, minimum width=2.2cm, minimum height=0.9cm, align=center, font=\small},
    compbox/.style={draw, rounded corners, fill=green!8, minimum width=2.0cm, minimum height=0.9cm, align=center, font=\small},
    outbox/.style={draw, rounded corners, fill=orange!12, minimum width=2.8cm, minimum height=0.9cm, align=center, font=\small},
    riskbox/.style={draw, rounded corners, fill=red!8, minimum width=2.4cm, minimum height=0.8cm, align=center, font=\small},
    arr/.style={-{Stealth[length=2.5mm]}, thick},
    node distance=0.6cm
]
\node[box] (onet1) {O*NET 30.2\\Tasks \& Ratings};
\node[box, right=1.0cm of onet1] (onet2) {O*NET 30.2\\Abilities};
\node[box, right=1.0cm of onet2] (bls) {BLS OEWS\\Employment};
\node[box, right=1.0cm of bls] (regional) {Regional\\Adoption Data};

\node[compbox, below=0.8cm of onet1] (wt) {$w_{o,t}$\\Task Weights};
\node[compbox, below=0.8cm of onet2] (cap) {CAP$(t)$\\AI Capability};
\node[compbox, below=0.8cm of bls] (cov) {COV$(t,o)$\\Coverage};
\node[compbox, below=0.8cm of regional] (v) {$V(t,r,\tau)$\\S-Curve};

\node[outbox, below=1.0cm of $(cap)!0.5!(cov)$] (ate) {$\mathrm{ATE}_o(r,\tau) = \sum w_{o,t} \cdot \mathrm{CAP}(t) \cdot \mathrm{COV}(t,o) \cdot V(t,r,\tau)$};

\node[riskbox, below=0.8cm of ate] (risk) {Risk Classification:\\High $\geq 0.65$ ~|~ Moderate 0.35--0.65 ~|~ Low $<0.35$};

\draw[arr] (onet1) -- (wt);
\draw[arr] (onet2) -- (cap);
\draw[arr] (bls) -- (cov);
\draw[arr] (regional) -- (v);
\draw[arr] (wt) -- (ate);
\draw[arr] (cap) -- (ate);
\draw[arr] (cov) -- (ate);
\draw[arr] (v) -- (ate);
\draw[arr] (ate) -- (risk);
\end{tikzpicture}
\caption{ATE computation pipeline. O*NET task statements and ratings feed into task weights ($w_{o,t}$), AI capability scores (CAP), and workflow coverage factors (COV). Regional adoption data parameterize the S-curve velocity function $V(t, r, \tau)$. The four components are multiplied and summed across tasks to produce occupation-level ATE scores, which are then classified by displacement risk threshold.}
\label{fig:pipeline}
\end{figure}

Figure~\ref{fig:pipeline} illustrates the full ATE computation pipeline from data inputs through score aggregation. The components are:
\begin{itemize}
\item $w_{o,t} \in [0,1]$ is the importance-weighted relevance of task $t$ in occupation $o$, drawn from O*NET task importance ratings normalized to sum to unity within each occupation;
\item $\mathrm{CAP}(t) \in [0,1]$ is the AI capability score for task $t$, measuring the current demonstrated ability of agentic AI systems to perform task $t$ at a level meeting professional standards, estimated from AI benchmark literature and practitioner deployment reports;
\item $\mathrm{COV}(t,o) \in [0,1]$ is the workflow coverage factor, measuring the degree to which task $t$ can be completed by an agentic system operating within occupation $o$'s standard workflow context (i.e., without human handoff for coordination, context-setting, or exception handling that requires occupational judgment).
\end{itemize}

The key theoretical innovation in the ATE score relative to prior exposure measures is the $\mathrm{COV}(t,o)$ term. Prior measures such as AIOE \cite{felten2021occupational} and the Eloundou et al.\ GPT-exposure measure \cite{eloundou2023gpts} set $\mathrm{COV}(t,o) = 1$ implicitly, treating every task as independently completable by AI. The ATE score explicitly penalizes tasks that require human coordination of multi-task workflows, inter-task context transfer, or exception handling beyond agentic AI's current operational envelope. Because agentic systems handle workflow coordination internally, $\mathrm{COV}(t,o)$ scores are higher than they would be under task-level AI analysis. This means ATE scores are closer to (and for well-bounded workflows, exceed) prior exposure estimates, and displacement risk is correspondingly higher for occupations with well-defined, digital workflows.

\subsection{Adoption Velocity and Temporal Projection}

The adoption velocity parameter $V(t, r, \tau)$ in Equation~(\ref{eq:ate}) captures the rate at which agentic AI systems capable of performing task $t$ will be deployed in region $r$ by time $\tau$. In the current implementation, $V$ does not vary across tasks within a region-year, so we simplify to $V(r, \tau) = L/(1+e^{-k(\tau-\tau_0)})$, a logistic (S-curve) adoption function \cite{bass1969new} where $\tau$ is the projection year and $\tau_0$ is the inflection point. The parameters $k$ (growth rate), $\tau_0$ (inflection point), and $L$ (saturation ceiling) are calibrated for each region using a three-tier metro classification adapted from the Brookings Institution's AI readiness taxonomy \cite{brookings2024ai}, and externally anchored against enterprise AI adoption surveys and employment outcome data (Section~\ref{sec:scurve_anchoring}). Table~\ref{tab:scurve} reports the calibrated values and their empirical basis.

\begin{table}[t]
\centering
\caption{S-Curve Adoption Parameters by Metro Tier}
\label{tab:scurve}
\small
\begin{tabular}{llcccp{5.5cm}}
\toprule
Region & Tier & $k$ & $\tau_0$ & $L$ & Calibration Basis \\
\midrule
SF Bay Area & 1 & 0.85 & 2024.25 & 0.92 & Highest AI patent density and VC investment per capita \cite{maslej2024ai} \\
Seattle & 2 & 0.62 & 2025.00 & 0.85 & AWS/Microsoft anchor employers; WARN Act data \cite{warn2024} \\
Austin & 2 & 0.62 & 2025.00 & 0.85 & Emerging hub; LinkedIn job posting trends \cite{linkedin2024future} \\
Boston & 2 & 0.62 & 2025.00 & 0.85 & University-anchored AI cluster \cite{maslej2024ai} \\
New York & 3 & 0.48 & 2025.75 & 0.78 & Diverse economy dilutes AI concentration \\
\bottomrule
\end{tabular}
\vspace{0.3em}

\noindent\footnotesize Tier classification adapts Brookings' AI readiness framework \cite{brookings2024ai}. The growth rate $k$ is bounded by McKinsey's global enterprise AI adoption rate ($k = 0.25$, lower bound) and Gartner's agentic AI enterprise app projection ($k \approx 2.1$, upper bound; see Section~\ref{sec:scurve_anchoring}); $\tau_0$ reflects each region's estimated inflection point consistent with McKinsey's finding that 62\% of enterprises were experimenting with AI agents by mid-2025 \cite{mckinsey2025ai}; $L$ reflects the proportion of occupations within agentic AI's operational envelope given regional industry composition. Sensitivity to these parameters is examined in Section~\ref{sec:sensitivity}; external anchoring against four independent data sources is in Section~\ref{sec:scurve_anchoring}.
\end{table}

An occupation is classified as \emph{high displacement risk} if $\mathrm{ATE}_o(r, \tau) \geq 0.65$, \emph{moderate risk} if $0.35 \leq \mathrm{ATE}_o(r, \tau) < 0.65$, and \emph{low risk} if $\mathrm{ATE}_o(r, \tau) < 0.35$. These thresholds are calibrated to correspond approximately to the 70\%, 35\%, and 10\% task displacement levels used in prior literature \cite{eloundou2023gpts,frey2017future}.

\section{Data and Methodology}
\label{sec:data}

\subsection{Occupational Task Data}

Task-level data for 236 Standard Occupational Classification (SOC) codes spanning six major groups (SOC~13 Financial, SOC~23 Legal, SOC~29 Healthcare, SOC~31 Healthcare Support, SOC~41 Sales, and SOC~43 Administrative/Clerical) are drawn from the O*NET 30.2 database \cite{onet2025}, which provides task statements, importance ratings (1--5 scale), and relevance ratings (0--100\% scale) for each occupation. Task importance weights $w_{o,t}$ are computed as the normalized product of importance and relevance ratings. After filtering to these six SOC groups, the analysis dataset contains 4,577 task statements (from the full O*NET corpus of 18,796 statements across all occupations).

\subsection{AI Capability Scores}

AI capability scores $\mathrm{CAP}(t)$ are estimated through a structured synthesis of: (1)~published AI benchmark results on tasks corresponding to occupation-relevant domains (MMLU, HumanEval, LexGLUE, MedQA, FinBen); (2)~published agentic AI evaluation results including SWE-bench for software engineering and AgentBench for general agentic capability; and (3)~practitioner reports from enterprise AI deployments in financial services, legal, healthcare, and customer operations published by McKinsey Global Institute \cite{mckinsey2023generative}, Goldman Sachs Research \cite{goldman2023ai}, and the Bureau of Labor Statistics \cite{bls2025ai}. CAP scores are operationalized using a BAIOE-style ability-weighted mapping: each occupation's O*NET ability profile (52 abilities with importance ratings) is combined with AI capability estimates for each ability category, derived from published benchmarks. Table~\ref{tab:cap} reports representative mappings. The occupation-level CAP for a task is the importance-weighted mean of AI capability scores across the occupation's ability profile, adjusted by a task-text modifier that boosts scores for tasks with strong AI-benchmark alignment (e.g., data entry, correspondence) and reduces them for tasks requiring physical or interpersonal execution.

\begin{table}[t]
\centering
\caption{Representative CAP Ability-to-AI Mappings (Selected Abilities)}
\label{tab:cap}
\small
\begin{tabular}{llcl}
\toprule
O*NET Ability & Category & AI Score & Benchmark Source \\
\midrule
Written Comprehension & Cognitive & 0.95 & MMLU $\sim$87\% \\
Written Expression & Cognitive & 0.93 & GPT-4 bar exam, CFA \\
Deductive Reasoning & Cognitive & 0.88 & MMLU Logic $\sim$88\% \\
Information Ordering & Cognitive & 0.85 & AgentBench planning \\
Mathematical Reasoning & Cognitive & 0.75 & MATH benchmark $\sim$75\% \\
Memorization & Cognitive & 0.60 & Context-window limited \\
Speed of Closure & Cognitive & 0.55 & Ambiguity resolution \\
Speech Recognition & Sensory & 0.45 & Whisper ASR WER $\sim$5--15\% \\
Near Vision & Sensory & 0.30 & Vision models (partial) \\
Manual Dexterity & Psychomotor & 0.10 & No physical capability \\
Static Strength & Physical & 0.00 & No physical capability \\
\bottomrule
\end{tabular}
\vspace{0.3em}

\noindent\footnotesize Full 52-ability mapping available in supplementary materials.
\end{table}

\subsection{Workflow Coverage Estimation}
\label{sec:cov}

Workflow coverage factors $\mathrm{COV}(t,o)$ are estimated using a structured rubric evaluating each task in the context of its occupation's standard workflow. Coverage is penalized when a task triggers one or more of four penalty categories, each reducing the base COV multiplicatively:
\begin{itemize}
\item \textbf{P1} ($-25\%$): \emph{Interpersonal context}. The task requires real-time rapport, negotiation, counseling, or emotional engagement not documentable in a task specification (keywords: negotiate, counsel, mediate, comfort, de-escalate, patient interaction);
\item \textbf{P2} ($-30\%$): \emph{Regulatory/fiduciary accountability}. The task involves legal liability, certification, sworn testimony, diagnosis, or binding agreements that current AI governance frameworks do not support for autonomous action (keywords: fiduciary, regulatory compliance, certify, notarize, prescribe, diagnose);
\item \textbf{P3} ($-40\%$): \emph{Physical presence}. The task requires sensorimotor coordination, on-site work, or manual handling (keywords: physically, lift, operate machinery, on-site, field work);
\item \textbf{P4} ($-20\%$): \emph{Exception handling}. The task involves crisis response, ambiguous situations, or novel judgments outside the standard envelope more than 30\% of the time (keywords: emergency, crisis, escalate, override, judgment call, novel situation).
\end{itemize}

\textbf{Worked examples from O*NET task statements.}  For Sales Representatives of Services (41-3091), the O*NET task ``Negotiate prices or terms of sales or service agreements'' triggers P1 (negotiate), yielding $\mathrm{COV} = 1.0 \times 0.75 = 0.75$. The task ``Maintain customer records using automated systems'' triggers no penalties ($\mathrm{COV} = 1.0$). Across all 15 O*NET tasks for this occupation, only 1 triggers a penalty, producing a mean COV of 0.98.

For Health Information Technologists (29-9021), the task ``Assign the patient to diagnosis-related groups (DRGs), using appropriate computer software'' triggers P2 (diagnos), yielding $\mathrm{COV} = 0.70$. The task ``Resolve or clarify codes or diagnoses with conflicting, missing, or unclear information'' also triggers P2. Of 16 tasks, 2 trigger penalties, producing a mean COV of 0.96.

\textbf{Limitation: keyword false negatives.}  The keyword rubric produces false negatives when interpersonal, regulatory, or physical dimensions are implicit rather than lexically present in the O*NET task text. For instance, ``Consult with clients after sales or contract signings to resolve problems'' is arguably interpersonal (P1) but does not contain the trigger keyword. The rubric is therefore a lower bound on penalties (upper bound on COV), and a validated expert rubric would be expected to reduce COV scores for mid-range occupations.

\textbf{Semantic coverage pilot.} To quantify the false-negative rate, we conducted a pilot evaluation using an expanded semantic pattern set on all 4,577 O*NET task statements across the 236 occupations in our study. The semantic patterns extend the keyword rubric with contextual triggers. For example, they add ``consult,'' ``confer,'' ``advise,'' and ``collaborate'' to the P1 interpersonal category, and ``approve,'' ``authorize,'' ``sentence,'' and ``verify compliance'' to P2 regulatory, while preserving the same four-category penalty structure.

Results indicate that the keyword rubric flags 632 tasks (13.8\%) with at least one penalty, while the semantic extension flags 1,142 tasks (25.0

The pilot reveals concrete implications for our top-ranked occupations. Credit Analysts (13-2041), ranked \#1, receive zero keyword penalties across all 11 tasks but would receive semantic penalties on 4 of 11 tasks: ``Consult with customers to resolve complaints'' (P1), ``Contact customers to collect payments on delinquent accounts'' (P1), ``Confer with credit association and other business representatives'' (P1), and ``Complete loan applications, including credit analyses and submit for approval'' (P2). Judges (23-1023), ranked \#2, would gain seven additional penalties (five P1, two P2) beyond the single keyword-detected penalty, including ``Sentence defendants in criminal cases'' (P2) and ``Preside over hearings and listen to allegations'' (P1).

These results confirm that the keyword rubric's ATE scores represent upper-bound displacement estimates. We retain the keyword rubric for the primary analysis due to its transparency and reproducibility, but recommend that future implementations: (1)~deploy full LLM-based semantic classification with multiple independent passes to compute inter-classifier agreement; (2)~report COV scores with confidence intervals derived from classifier variance; and (3)~validate against expert human ratings on a stratified sample of at least 200 tasks spanning all six SOC groups. The complete keyword lists, semantic patterns, and pilot results are available from the corresponding author upon request.

\subsection{Regional Employment and Adoption Data}

Employment counts by occupation and region are drawn from the BLS Occupational Employment and Wage Statistics (OEWS) program \cite{bls2024oews}, using the May 2024 estimates for the five Metropolitan Statistical Areas under study. Regional adoption velocity parameters are calibrated using: Washington State Employment Security Department WARN Act notices \cite{warn2024}, LinkedIn Economic Graph job posting trend data \cite{linkedin2024future}, and enterprise AI investment reports for each region.

The five regions and their 2024 total employment (thousands) are: Seattle--Tacoma--Bellevue (1,847); San Francisco Bay Area (2,431); Austin--Round Rock (1,234); New York--Newark (9,872); and Boston--Cambridge (2,298). Combined, these MSAs account for approximately 11.3\% of total US non-farm employment.

\subsection{Remote Work and the Erosion of Geographic Adoption Buffers}
\label{sec:remotework}

A structural limitation of purely geography-based adoption models is that remote and hybrid work arrangements decouple a worker's physical location from their employer's technology adoption tier. A financial analyst residing in Austin (Tier~2) but employed by a San Francisco--headquartered firm (Tier~1) faces the adoption velocity of the employer, not the residence metro. The BLS Current Population Survey publishes telework prevalence data by occupation \cite{bls2024telework}, and the information-intensive SOC groups in our study population exhibit sharply divergent telework rates, as shown in Table~\ref{tab:telework}.

\begin{table}[ht]
\centering
\caption{BLS CPS Table 60: Telework Rates by Occupation Group (2024 Annual Averages)}
\label{tab:telework}
\small
\begin{tabular}{lrrrrr}
\toprule
\textbf{SOC Group} & \textbf{Workers (K)} & \textbf{Total \%} & \textbf{Some Hrs \%} & \textbf{All Hrs \%} & $\boldsymbol{R_o}$ \\
\midrule
Business \& Financial Ops (13) & 9{,}876 & 55.9 & 28.1 & 27.9 & 0.559 \\
Legal (23) & 1{,}822 & 52.7 & 33.2 & 19.5 & 0.527 \\
Healthcare Practitioners (29) & 9{,}974 & 13.1 & 8.1 & 5.0 & 0.131 \\
Healthcare Support (31) & 5{,}620 & 10.2 & 2.6 & 7.6 & 0.102 \\
Sales \& Related (41) & 13{,}710 & 23.8 & 12.3 & 11.5 & 0.238 \\
Office \& Admin Support (43) & 15{,}920 & 25.1 & 11.1 & 14.0 & 0.251 \\
\bottomrule
\end{tabular}
\vspace{0.3em}

\noindent\footnotesize Source: BLS Current Population Survey, 2024 Annual Averages, Table 60. ``Total \%'' is the share of workers who teleworked or worked at home for pay. $R_o$ is the telework rate expressed as a proportion for use in Equation~\ref{eq:veff}.
\end{table}

To account for this, we introduce a remote work feasibility index $R_o \in [0,1]$ that adjusts the effective adoption velocity experienced by workers in a given occupation. The effective adoption velocity blends the residence region's $V$ with an employer-tier-weighted $V$:
\begin{equation}
V_{\mathrm{eff}}(o, r, \tau) = (1 - R_o) \cdot V(r, \tau) + R_o \cdot \sum_{j} \pi_j \cdot V(\mathrm{tier}_j, \tau)
\label{eq:veff}
\end{equation}
where $R_o$ is the BLS-reported telework rate for occupation $o$ (Table~\ref{tab:telework}), and $\pi_j$ is the share of industry employment in adoption tier $j$, computed from BLS QCEW 2024 Annual Averages by Industry \cite{bls2024qcew}. This formulation avoids the need for individual employer--residence microdata: instead of tracking which specific employer each remote worker belongs to, we use the observed employment distribution across our five study MSAs as a proxy for the employer-tier mix.

\begin{table}[ht]
\centering
\caption{Employer Tier Distribution $\pi_j$ from QCEW 2024 and Remote Work Adjustment Impact (2027)}
\label{tab:veff}
\small
\begin{tabular}{l rrr rrr}
\toprule
& \multicolumn{3}{c}{\textbf{Employer Tier Share $\pi_j$}} & \multicolumn{3}{c}{\textbf{Mean ATE $\Delta$ (2027)}} \\
\cmidrule(lr){2-4} \cmidrule(lr){5-7}
\textbf{SOC Group} & Tier~1 & Tier~2 & Tier~3 & SF Bay & Tier~2 & NY \\
\midrule
Financial (13) & 12.2\% & 27.1\% & 60.7\% & $-$16.8\% & $-$6.1\% & $+$9.2\% \\
Legal (23) & 14.6\% & 23.2\% & 62.1\% & $-$15.7\% & $-$5.6\% & $+$8.9\% \\
HC Practitioners (29) & 15.9\% & 26.7\% & 57.4\% & $-$3.8\% & $-$1.2\% & $+$2.5\% \\
HC Support (31) & 15.9\% & 26.7\% & 57.4\% & $-$2.9\% & $-$0.9\% & $+$1.9\% \\
Sales (41) & 15.9\% & 33.3\% & 50.7\% & $-$6.5\% & $-$1.8\% & $+$5.0\% \\
Admin/Clerical (43) & 10.2\% & 33.8\% & 56.0\% & $-$7.4\% & $-$2.6\% & $+$4.3\% \\
\bottomrule
\end{tabular}
\vspace{0.3em}

\noindent\footnotesize Left panel: employer tier distribution computed from QCEW 2024 private-sector employment across the five study MSAs. Tier~1 = SF Bay Area, Tier~2 = Seattle/Austin/Boston, Tier~3 = New York. Right panel: percentage change in mean ATE scores under the remote-adjusted model (Equation~\ref{eq:veff}) relative to the residence-based baseline, by region and SOC group.
\end{table}

Table~\ref{tab:veff} reveals that remote work produces a \emph{convergence} effect rather than the unidirectional geographic buffer erosion hypothesized in prior literature. The QCEW data show that only 12--16\% of industry employment in our study MSAs is located in Tier~1 (SF Bay Area), while 50--62\% is in Tier~3 (New York). Consequently, the employer-weighted adoption velocity $\sum_j \pi_j V(\mathrm{tier}_j, \tau)$ is \emph{lower} than SF Bay's local $V$ but \emph{higher} than New York's local $V$.

For the San Francisco Bay Area (Tier~1), the remote-adjusted model reduces mean Financial ATE by 16.8\% and Legal ATE by 15.7\% in 2027, because remote workers in SF Bay are employed by firms distributed across all tiers, not exclusively local Tier~1 firms. This means the residence-based model \emph{overestimates} displacement risk for Tier~1 regions. For New York (Tier~3), the adjustment increases Financial ATE by 9.2\% and Legal ATE by 8.9\%, confirming that remote work transmits higher-tier adoption pressure into slower-adopting regions. Tier~2 regions show modest decreases ($-$5--6\% for Financial and Legal). Healthcare occupations are minimally affected across all regions ($\pm$1--4\%) due to their low telework rates ($R_o < 0.14$).

The net effect is regional convergence: remote work compresses the ATE gap between the fastest and slowest-adopting metros. We retain the residence-based $V(r,\tau)$ as the primary specification in Tables~\ref{tab:top20} and~\ref{tab:regional}, and report the remote-adjusted results as a sensitivity analysis. The full remote-adjusted ATE scores for all 236 occupations are available from the corresponding author upon request.

\section{Results}
\label{sec:results}

\subsection{ATE Score Distribution Across Occupations}

\begin{figure}[t]
\centering
\begin{tikzpicture}
\def\xsc{28}  
\def\ysc{0.032}  
\draw[->] ({0.13*\xsc},0) -- ({0.48*\xsc},0) node[right] {\small ATE Score};
\draw[->] ({0.14*\xsc},{-5*\ysc}) -- ({0.14*\xsc},{210*\ysc}) node[above] {\small Count};
\foreach \y in {0,50,100,150,200} {
  \draw ({0.14*\xsc},{\y*\ysc}) -- ({0.137*\xsc},{\y*\ysc}) node[left, font=\footnotesize] {\y};
}
\foreach \x/\l in {0.15/0.15, 0.20/0.20, 0.25/0.25, 0.30/0.30, 0.35/0.35, 0.40/0.40, 0.45/0.45} {
  \draw ({\x*\xsc},{-2*\ysc}) -- ({\x*\xsc},0);
  \node[below, font=\footnotesize] at ({\x*\xsc},{-3*\ysc}) {\l};
}
\draw[dashed, red!70!black, thick] ({0.35*\xsc},{-5*\ysc}) -- ({0.35*\xsc},{210*\ysc});
\node[red!70!black, font=\scriptsize, rotate=90, anchor=south] at ({0.348*\xsc},{140*\ysc}) {ATE $= 0.35$};

\fill[orange!60, opacity=0.6] ({0.15*\xsc},0) rectangle ({0.20*\xsc},{38*\ysc});
\fill[orange!60, opacity=0.6] ({0.20*\xsc},0) rectangle ({0.25*\xsc},{177*\ysc});
\fill[orange!60, opacity=0.6] ({0.25*\xsc},0) rectangle ({0.30*\xsc},{21*\ysc});
\draw[orange!80!black, thin] ({0.15*\xsc},0) rectangle ({0.20*\xsc},{38*\ysc});
\draw[orange!80!black, thin] ({0.20*\xsc},0) rectangle ({0.25*\xsc},{177*\ysc});
\draw[orange!80!black, thin] ({0.25*\xsc},0) rectangle ({0.30*\xsc},{21*\ysc});

\fill[green!45!black, opacity=0.45] ({0.15*\xsc},0) rectangle ({0.20*\xsc},{1*\ysc});
\fill[green!45!black, opacity=0.45] ({0.20*\xsc},0) rectangle ({0.25*\xsc},{13*\ysc});
\fill[green!45!black, opacity=0.45] ({0.25*\xsc},0) rectangle ({0.30*\xsc},{121*\ysc});
\fill[green!45!black, opacity=0.45] ({0.30*\xsc},0) rectangle ({0.35*\xsc},{101*\ysc});
\draw[green!60!black, thin] ({0.25*\xsc},0) rectangle ({0.30*\xsc},{121*\ysc});
\draw[green!60!black, thin] ({0.30*\xsc},0) rectangle ({0.35*\xsc},{101*\ysc});

\fill[blue!55, opacity=0.7] ({0.25*\xsc},0) rectangle ({0.30*\xsc},{4*\ysc});
\fill[blue!55, opacity=0.7] ({0.30*\xsc},0) rectangle ({0.35*\xsc},{65*\ysc});
\fill[blue!55, opacity=0.7] ({0.35*\xsc},0) rectangle ({0.40*\xsc},{99*\ysc});
\fill[blue!55, opacity=0.7] ({0.40*\xsc},0) rectangle ({0.45*\xsc},{68*\ysc});
\draw[blue!70!black, thin] ({0.35*\xsc},0) rectangle ({0.40*\xsc},{99*\ysc});
\draw[blue!70!black, thin] ({0.40*\xsc},0) rectangle ({0.45*\xsc},{68*\ysc});
\draw[blue!70!black, thin] ({0.30*\xsc},0) rectangle ({0.35*\xsc},{65*\ysc});

\draw[gray!50, thin, fill=white, opacity=0.9] ({0.375*\xsc},{165*\ysc}) rectangle ({0.465*\xsc},{205*\ysc});
\fill[blue!55, opacity=0.7] ({0.38*\xsc},{192*\ysc}) rectangle ({0.395*\xsc},{200*\ysc});
\draw[blue!70!black, thin] ({0.38*\xsc},{192*\ysc}) rectangle ({0.395*\xsc},{200*\ysc});
\node[right, font=\scriptsize] at ({0.398*\xsc},{196*\ysc}) {SF Bay (Tier~1)};
\fill[green!45!black, opacity=0.45] ({0.38*\xsc},{179*\ysc}) rectangle ({0.395*\xsc},{187*\ysc});
\draw[green!60!black, thin] ({0.38*\xsc},{179*\ysc}) rectangle ({0.395*\xsc},{187*\ysc});
\node[right, font=\scriptsize] at ({0.398*\xsc},{183*\ysc}) {Seattle (Tier~2)};
\fill[orange!60, opacity=0.6] ({0.38*\xsc},{166*\ysc}) rectangle ({0.395*\xsc},{174*\ysc});
\draw[orange!80!black, thin] ({0.38*\xsc},{166*\ysc}) rectangle ({0.395*\xsc},{174*\ysc});
\node[right, font=\scriptsize] at ({0.398*\xsc},{170*\ysc}) {New York (Tier~3)};
\end{tikzpicture}
\caption{Distribution of 2027 ATE scores across all 236 occupations for three representative regions. The dashed red line marks the moderate-risk threshold ($\mathrm{ATE} \geq 0.35$). In the SF Bay Area (Tier~1), 70.8\% of occupations exceed this threshold; in Seattle (Tier~2), 0\%; in New York (Tier~3), 0\%. The rightward shift from Tier~3 to Tier~1 reflects the adoption velocity differential parameterized by the S-curve model.}
\label{fig:ate_dist}
\end{figure}
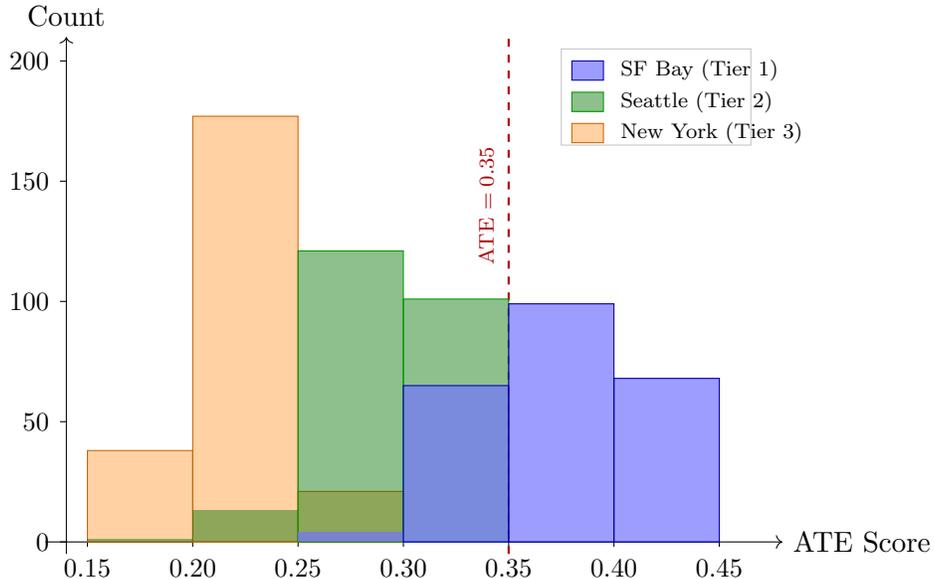

Table~\ref{tab:top20} presents the twenty occupations with the highest projected ATE scores by 2027 in the San Francisco Bay Area (Tier~1), selected as the upper-bound adoption scenario. Figure~\ref{fig:ate_dist} shows the full distribution of 2027 ATE scores across all 236 occupations for three representative regions. In the SF Bay Area (Tier~1), the distribution is concentrated in the moderate-risk range: 99 occupations (41.9\%) fall in the 0.35--0.40 band and 68 (28.8\%) in the 0.40--0.45 band, with only 69 occupations (29.2\%) remaining below the moderate-risk threshold. The distribution shifts leftward with decreasing adoption tier: Tier~2 (Seattle) is concentrated in the 0.25--0.30 modal bin, while Tier~3 (New York) centers on 0.20--0.25. This tier-dependent shift illustrates the adoption velocity effect: the same 236 occupations exhibit markedly different risk profiles depending on regional deployment speed.

\begin{table}[t]
\centering
\caption{Top 20 Occupations by Projected ATE Score (San Francisco Bay Area), 2027}
\label{tab:top20}
\small
\begin{tabular}{lp{2.4cm}ccc}
\toprule
Occupation (SOC) & Category & 2025 & 2027 & 2030 \\
\midrule
Credit Analysts (13-2041) & Financial & 0.31 & 0.43 & 0.47 \\
Judges \& Magistrates (23-1023)$^\ddagger$ & Legal & 0.31 & 0.43 & 0.47 \\
Sustainability Specialists (13-1199) & Financial & 0.31 & 0.43 & 0.47 \\
Regulatory Affairs Specialists (13-1041.07) & Financial & 0.31 & 0.43 & 0.47 \\
Market Research Analysts (13-1161) & Financial & 0.31 & 0.43 & 0.47 \\
Preventive Medicine Physicians (29-1229)$^\dagger$ & Healthcare & 0.31 & 0.43 & 0.47 \\
Business Continuity Planners (13-1199) & Financial & 0.30 & 0.42 & 0.46 \\
Financial Examiners (13-2061) & Financial & 0.30 & 0.42 & 0.46 \\
Statistical Assistants (43-9111) & Admin/Clerical & 0.30 & 0.42 & 0.46 \\
Search Marketing Strategists (13-1161) & Financial & 0.30 & 0.42 & 0.46 \\
Cost Estimators (13-1051) & Financial & 0.30 & 0.42 & 0.46 \\
Equal Opportunity Representatives (13-1041.03) & Financial & 0.30 & 0.42 & 0.46 \\
Labor Relations Specialists (13-1075) & Financial & 0.30 & 0.42 & 0.46 \\
Financial Risk Specialists (13-2054) & Financial & 0.30 & 0.42 & 0.46 \\
First-Line Supervisors (41-1012) & Sales & 0.30 & 0.42 & 0.46 \\
Securities Sales Agents (41-3031) & Sales & 0.30 & 0.42 & 0.46 \\
Insurance Underwriters (13-2053) & Financial & 0.30 & 0.42 & 0.46 \\
Personal Financial Advisors (13-2052) & Financial & 0.30 & 0.42 & 0.46 \\
Admin. Law Judges (23-1021) & Legal & 0.30 & 0.42 & 0.46 \\
Logisticians (13-1081) & Financial & 0.30 & 0.42 & 0.45 \\
\bottomrule
\end{tabular}
\vspace{0.3em}

\noindent\footnotesize $^\dagger$Healthcare practitioner occupations whose ATE scores may overstate displacement risk due to the keyword-based COV rubric's known false-negative limitation for implicit physical presence requirements (see Section~\ref{sec:cov}). $^\ddagger$Judicial occupations score high on technical AI capability for legal research, document analysis, and case management tasks, but face strong institutional and constitutional protections (Article~III tenure, separation of powers) that the ATE framework does not model; the score reflects \emph{technical exposure} to agentic AI capabilities, not practical displacement likelihood for constitutionally protected roles.
\end{table}

Several observations emerge from Table~\ref{tab:top20}, which reports results for the San Francisco Bay Area (Tier~1 adoption region) to illustrate the upper bound of near-term displacement dynamics. First, financial and legal occupations rank highest, reflecting their combination of high AI capability scores and well-bounded digital workflows. Credit Analysts (13-2041) lead at ATE 0.47 in 2030, tied with Judges and Magistrates (23-1023) and Sustainability Specialists (13-1199). Second, the growth from 2025 to 2030 scores (e.g., Credit Analysts: 0.31 to 0.47) is driven by the S-curve adoption model: the inflection point for Tier~1 regions is calibrated at Q2~2024, meaning deployment has already passed the steepest acceleration phase and is approaching saturation. Third, occupations ranked 1--20 in Table~\ref{tab:top20} cluster tightly in the 0.42--0.47 ATE range for 2027--2030, reflecting the workflow coverage penalties applied to tasks requiring interpersonal coordination and exception handling that agentic systems do not yet complete autonomously. Fourth, the preventive medicine physician entry (29-1229, marked $\dagger$ in the table) appears despite requiring physical presence; this reflects the keyword-based COV rubric's known false-negative limitation for implicit physical requirements (see Section~\ref{sec:cov}). A notable feature of Table~\ref{tab:top20} is that all top-20 occupations share ATE scores within a narrow band (0.42--0.47 by 2027), and no occupation in our study reaches the high-risk threshold ($\mathrm{ATE} \geq 0.65$) even by 2030. This reflects genuine similarity in their underlying ability profiles and the effect of the COV penalty structure: these occupations are predominantly cognitive-digital with few physical or interpersonal penalty triggers, producing convergent ATE estimates, but each retains sufficient exception-handling, regulatory, or interpersonal task components to prevent scores from reaching the upper range. The implication is important: our framework predicts \emph{widespread moderate risk} across information-processing occupations rather than catastrophic displacement of any single occupation. The displacement pressure is broad but bounded, consistent with the gradual workforce recomposition pattern rather than mass layoff scenarios. The framework differentiates meaningfully between these moderate-exposure information-processing roles and low-exposure occupations requiring substantial physical or interpersonal interaction (which score below 0.35).

\subsection{Regional Variation}

Table~\ref{tab:regional} shows the share of occupations within each major category that cross the moderate-risk threshold ($\mathrm{ATE}_{2027} \geq 0.35$) across our five study regions, computed from our full 236-occupation dataset.

\begin{table}[t]
\centering
\caption{Share of Occupations Crossing ATE $\geq 0.35$ (\%) and Mean ATE by Region, 2027 vs.\ 2030}
\label{tab:regional}
\small
\begin{tabular}{l cc cc cc cc cc}
\toprule
& \multicolumn{2}{c}{Seattle} & \multicolumn{2}{c}{SF Bay} & \multicolumn{2}{c}{Austin} & \multicolumn{2}{c}{New York} & \multicolumn{2}{c}{Boston} \\
\cmidrule(lr){2-3} \cmidrule(lr){4-5} \cmidrule(lr){6-7} \cmidrule(lr){8-9} \cmidrule(lr){10-11}
Category & '27 & '30 & '27 & '30 & '27 & '30 & '27 & '30 & '27 & '30 \\
\midrule
Admin/Clerical (43) & 0.0 & 74.5 & 78.4 & 92.2 & 0.0 & 74.5 & 0.0 & 0.0 & 0.0 & 74.5 \\
Financial (13) & 0.0 & 87.5 & 91.7 & 95.8 & 0.0 & 87.5 & 0.0 & 8.3 & 0.0 & 87.5 \\
Legal (23) & 0.0 & 100.0 & 100.0 & 100.0 & 0.0 & 100.0 & 0.0 & 14.3 & 0.0 & 100.0 \\
Healthcare (29) & 0.0 & 43.8 & 65.2 & 98.9 & 0.0 & 43.8 & 0.0 & 1.1 & 0.0 & 43.8 \\
Sales (41) & 0.0 & 63.6 & 68.2 & 95.5 & 0.0 & 63.6 & 0.0 & 0.0 & 0.0 & 63.6 \\
HC Support (31) & 0.0 & 10.5 & 15.8 & 57.9 & 0.0 & 10.5 & 0.0 & 0.0 & 0.0 & 10.5 \\
\midrule
\multicolumn{11}{l}{\textit{Mean ATE by Category (2027 / 2030)}} \\
\midrule
Admin/Clerical (43) & \multicolumn{2}{c}{.30 / .37} & \multicolumn{2}{c}{.38 / .41} & \multicolumn{2}{c}{.30 / .37} & \multicolumn{2}{c}{.23 / .31} & \multicolumn{2}{c}{.30 / .37} \\
Financial (13) & \multicolumn{2}{c}{.31 / .39} & \multicolumn{2}{c}{.40 / .43} & \multicolumn{2}{c}{.31 / .39} & \multicolumn{2}{c}{.24 / .33} & \multicolumn{2}{c}{.31 / .39} \\
Legal (23) & \multicolumn{2}{c}{.33 / .40} & \multicolumn{2}{c}{.42 / .45} & \multicolumn{2}{c}{.33 / .40} & \multicolumn{2}{c}{.25 / .34} & \multicolumn{2}{c}{.33 / .40} \\
Healthcare (29) & \multicolumn{2}{c}{.28 / .35} & \multicolumn{2}{c}{.36 / .39} & \multicolumn{2}{c}{.28 / .35} & \multicolumn{2}{c}{.22 / .30} & \multicolumn{2}{c}{.28 / .35} \\
Sales (41) & \multicolumn{2}{c}{.29 / .36} & \multicolumn{2}{c}{.38 / .41} & \multicolumn{2}{c}{.29 / .36} & \multicolumn{2}{c}{.23 / .31} & \multicolumn{2}{c}{.29 / .36} \\
HC Support (31) & \multicolumn{2}{c}{.26 / .32} & \multicolumn{2}{c}{.33 / .35} & \multicolumn{2}{c}{.26 / .32} & \multicolumn{2}{c}{.20 / .27} & \multicolumn{2}{c}{.26 / .32} \\
\bottomrule
\end{tabular}
\vspace{0.3em}

\noindent\footnotesize Upper panel: percentage of occupations in each SOC group crossing $\mathrm{ATE} \geq 0.35$. Lower panel: category-level mean ATE (2027~/ 2030). The 2027-to-2030 comparison reveals temporal convergence: Tier~2 regions reach Tier~1's 2027 displacement levels by 2030.
\end{table}

The regional pattern in Table~\ref{tab:regional} reveals a striking concentration effect driven by adoption velocity. At the $\mathrm{ATE}_{2027} \geq 0.35$ threshold, the San Francisco Bay Area (Tier~1) shows substantially higher displacement rates across all occupational categories: 78.4\% of Administrative/Clerical, 91.7\% of Financial, and 100\% of Legal occupations cross the moderate-risk threshold. Tier~2 regions (Seattle, Austin, Boston) show zero crossings at the same 2027 horizon, but the mean ATE values reveal a more detailed picture than the binary threshold suggests: Tier~2 Financial occupations average 0.31 in 2027, just 0.04 below the 0.35 cutoff, indicating these regions sit at the threshold boundary rather than operating in a different regime.

The 2030 columns demonstrate temporal convergence. In aggregate, 93.2\% of all 236 analyzed occupations (220 of 236) cross the moderate-risk threshold in the SF Bay Area by 2030, with only Healthcare Support (57.9\%) remaining substantially below saturation. By 2030, Tier~2 regions reach displacement levels comparable to SF Bay Area's 2027 position: 87.5\% of Financial and 100\% of Legal occupations cross the threshold in Seattle, Austin, and Boston by 2030, closely matching SF Bay's 91.7\% and 100\% in 2027. This 2--3 year lag is the central empirical finding: the same occupations face offset displacement timelines depending on local adoption velocity, and the offset is quantifiable from the S-curve parameters. Tier~3 (New York) lags further, with only 8.3\% of Financial and 14.3\% of Legal occupations crossing by 2030, reflecting its lower saturation ceiling ($L = 0.78$) and later inflection point.

The policy implication is that workforce transition programs must be regionally staged: Tier~1 interventions are needed now, Tier~2 interventions should be planned and budgeted now for deployment in 2028--2029, and Tier~3 regions have a wider but not indefinite planning window.

\subsection{Comparison with Prior Estimates}

Our ATE framework produces estimates that are more conservative than raw GPT-4 token-level exposure analysis but more actionable for labor market policy, because ATE scores represent effective deployment probability rather than theoretical capability. Eloundou et al.\ \cite{eloundou2023gpts} found that GPT-4 could affect at least 10\% of tasks for 80\% of the U.S.\ workforce; our ATE estimates capture a narrower but more consequential subset: occupations where agentic AI can complete workflows end-to-end without human handoff at current deployment scale.

For bookkeeping and accounting clerks (43-3031), the highest-frequency occupation in prior literature (Frey and Osborne estimated 98\% computerization probability; Arntz et al.\ revised this to 35\% using task-level analysis), our computed ATE in the San Francisco Bay Area by 2027 is 0.41, reflecting that agentic bookkeeping systems are actively deployed at enterprise scale in Tier~1 regions but that audit review and exception handling still require human oversight for compliance reasons. Notably, our highest-ATE occupations by 2027 are credit analysts, judges, and financial specialists (13-2041, 23-1023, ATE 0.43). These occupations have high cognitive AI-capability scores and well-bounded digital workflows. This finding diverges from prior task-level analyses that concentrated displacement risk in routine clerical work, and suggests that agentic AI's workflow-completion capability shifts the frontier of displacement toward information-intensive professional roles.

\subsection{Sensitivity Analysis}
\label{sec:sensitivity}

To assess the stability of our results under parameter uncertainty, we conduct one-at-a-time sensitivity analysis on the three S-curve parameters and the COV penalty weights, using the San Francisco Bay Area (Tier~1) as the baseline.

\textbf{S-curve parameters.}  The saturation ceiling $L$ exerts the largest influence: a $\pm 10\%$ change in $L$ produces a $\pm 10.0\%$ change in $V(2027)$. Growth rate $k$ has moderate impact: $\pm 20\%$ variation in $k$ shifts $V(2027)$ by $-5.0\%$ to $+3.4\%$. The inflection point $\tau_0$ shows symmetric sensitivity: shifting $\tau_0$ by $\pm 6$ months changes $V(2027)$ by approximately $\pm 3$--$5\%$. Critically, no parameter perturbation within plausible ranges changes the ordinal ranking of occupations or the finding that Tier~1 regions show dramatically higher displacement shares than Tier~2 and Tier~3 regions. Reclassifying Seattle from Tier~2 to Tier~1 would increase its projected adoption level $V(2027)$ by 27.3\%, but the regional divergence finding persists because the tier assignment reflects structural differences in AI employer density and venture capital concentration.

\textbf{Three-scenario $k$ stress test.}  The one-at-a-time analysis above perturbs $k$ by $\pm 20\%$; however, Section~\ref{sec:scurve_anchoring} documents a more substantial concern: the calibrated Tier~1 value ($k = 0.85$) exceeds the logistic fit to Indeed job-posting data ($k \approx 0.14$) by a factor of approximately~6. To assess whether this discrepancy undermines the paper's qualitative conclusions, we run a full three-scenario stress test spanning a range that brackets both values. Table~\ref{tab:k-sensitivity} reports the percentage of occupations crossing the moderate-risk threshold ($\mathrm{ATE} \geq 0.35$) for each tier under Conservative ($k_{\text{T1}}=0.40$, $k_{\text{T2}}=0.30$, $k_{\text{T3}}=0.23$), Baseline (current calibration), and Aggressive ($k_{\text{T1}}=1.20$, $k_{\text{T2}}=0.85$, $k_{\text{T3}}=0.65$) scenarios. All other parameters ($L$, $\tau_0$, occupation base-ATE components) are held fixed; only~$k$ varies.

\begin{table}[h!]
  \centering
  \caption{%
    \textbf{$k$-Sensitivity Analysis: Percentage of Occupations Crossing $\mathrm{ATE} \geq 0.35$ Risk Threshold.}
    Three adoption-velocity scenarios stress-test the S-curve steepness parameter~$k$,
    spanning a Conservative ($k_{\text{T1}}=0.40$) to Aggressive ($k_{\text{T1}}=1.20$) range.
    This range deliberately brackets the 6$\times$ discrepancy between the calibrated
    Baseline value ($k=0.85$) and the empirically fitted trend from job-posting
    data ($k\approx0.13$; Section~\ref{sec:scurve_anchoring}), assessing whether the qualitative risk ordering
    SF\,Bay\,$>$\,Tier\,2\,$>$\,New\,York holds regardless of $k$ assumption.}
  \label{tab:k-sensitivity}
  \setlength{\tabcolsep}{5pt}
  \begin{tabular}{l rrr rrr rrr}
    \toprule
    & \multicolumn{3}{c}{\textit{Conservative}} & \multicolumn{3}{c}{\textit{Baseline}} & \multicolumn{3}{c}{\textit{Aggressive}} \\
    \cmidrule(lr){2-4}\cmidrule(lr){5-7}\cmidrule(lr){8-10}
    \textbf{Tier / Metro} & \textbf{2025} & \textbf{2027} & \textbf{2030}
                          & \textbf{2025} & \textbf{2027} & \textbf{2030}
                          & \textbf{2025} & \textbf{2027} & \textbf{2030} \\
    \midrule
    \textbf{Tier 1} SF Bay Area
      & 0.0 &  2.5 & 69.9
      & 0.0 & 70.8 & 93.2
      & 0.0 & 87.3 & 94.1 \\
    \textbf{Tier 2} Seattle / Austin / Boston
      & 0.0 &  0.0 &  3.4
      & 0.0 &  0.0 & 60.2
      & 0.0 & 19.9 & 70.3 \\
    \textbf{Tier 3} New York
      & 0.0 &  0.0 &  0.0
      & 0.0 &  0.0 &  2.5
      & 0.0 &  0.0 & 28.8 \\
    \bottomrule
  \end{tabular}
  \begin{minipage}{\linewidth}
    \smallskip\footnotesize
    \textit{Note:} Values are the percentage of 236 O*NET occupations with recomputed $\mathrm{ATE} \geq 0.35$ under each scenario.
    Base ATE components ($w_{o,t}$, $\mathrm{CAP}$, $\mathrm{COV}$) are held fixed; only~$k$ varies.
    S-curve parameters: Tier~1 ($\tau_0=2024.25$, $L=0.92$);
    Tier~2 ($\tau_0=2025.00$, $L=0.85$);
    Tier~3 ($\tau_0=2025.75$, $L=0.78$).
  \end{minipage}
\end{table}

The qualitative risk ordering SF Bay $>$ Tier~2 $>$ New~York holds in all nine scenario-year combinations, confirming that the tier hierarchy is not an artifact of the assumed $k$ value. The magnitude and timing of threshold crossings are, however, highly $k$-sensitive: by 2027, SF Bay crosses the 0.35 threshold for only 2.5\% of occupations under Conservative assumptions but 87.3\% under Aggressive---a 35-fold difference---while New York remains at 0\% until 2030 under Conservative and Baseline scenarios and reaches only 28.8\% under the most rapid adoption scenario. These results confirm that while the \emph{timing and scale} of displacement risk depend substantially on the true steepness of the adoption curve (an open empirical question), the \emph{relative vulnerability} of technology-hub metros versus financial-center metros is a robust qualitative finding that persists across the full range of plausible $k$ values.

\textbf{COV penalty weights.}  Halving all four penalty weights (P1--P4) increases the highest-ATE occupation's score from 0.47 to approximately 0.48, a modest 2\% increase. This limited sensitivity reflects the fact that the highest-scoring occupations (credit analysts, judges, financial specialists) already have few penalty-triggering tasks. COV sensitivity is more pronounced for mid-range occupations: correspondence clerks (ATE 0.45) would rise to approximately 0.49 under halved penalties, potentially crossing the high-moderate boundary. This shows the importance of the COV rubric for occupations in the 0.35--0.65 range.

\textbf{CAP scores.} A uniform $\pm 10\%$ shift in all cognitive ability AI scores shifts ATE scores by approximately $\pm 7$--$8\%$, preserving relative rankings. Because CAP enters multiplicatively with task weights that are normalized within each occupation, absolute CAP changes affect all occupations proportionally, leaving the cross-occupation and cross-regional comparative analysis stable.

\subsection{External Validation of CAP Scores}
\label{sec:capvalidation}

To assess whether our AI capability scoring methodology captures a genuine exposure signal rather than arbitrary calibration, we compare ATE scores against two independently computed AI exposure indices using different methodologies.

\textbf{Convergent validity with AIOE.} Felten, Raj, and Seamans \cite{felten2021occupational} constructed the AI Occupational Exposure (AIOE) index by mapping documented AI capabilities to O*NET abilities. This methodology is structurally similar to our CAP scoring but was developed independently with different AI capability benchmarks and weighting schemes. Matching 193 of our 236 occupations by SOC code, we find a Spearman rank correlation of $\rho = 0.84$ ($p < 10^{-6}$) between mean ATE scores and AIOE scores. The strong correlation confirms that our CAP methodology captures the same underlying exposure signal as an independently validated measure.

\textbf{Convergent validity with GPT exposure.} Eloundou et al.\ \cite{eloundou2023gpts} estimated GPT-4 exposure scores using a different approach: combined human annotator and GPT-4 self-assessment ratings at the task level, aggregated to occupations. Matching all 236 occupations, we find $\rho = 0.72$ ($p < 10^{-6}$). The moderately lower correlation relative to AIOE is expected: the Eloundou et al.\ measure captures raw LLM capability exposure without workflow coverage adjustments, while our ATE score incorporates COV penalties that systematically reduce scores for occupations with interpersonal, regulatory, or physical barriers to agentic automation.

\textbf{Interpretable divergences.} Occupations where ATE ranks substantially lower than AIOE or GPT exposure (such as Postal Service Clerks, ATE rank 159 vs.\ AIOE rank 16, and Travel Agents, ATE rank 175 vs.\ AIOE rank 64) are precisely those with high COV penalty loads from physical presence (P3) or interpersonal engagement (P1) requirements. This pattern confirms that the COV term performs its intended function: discounting raw AI capability exposure for occupations where workflow barriers prevent end-to-end agentic automation.

\subsection{S-Curve Parameter Anchoring}
\label{sec:scurve_anchoring}

The S-curve parameters in Table~\ref{tab:scurve} are calibrated values rather than directly estimated from a single time series. To assess whether these parameters fall within empirically plausible ranges, we triangulate against four independent data sources spanning different facets of AI adoption.

\textbf{Growth rate ($k$) bounds.} Fitting a logistic function to the McKinsey Global AI Survey's enterprise adoption time series ($n = 9$ annual observations, 2017--2025) yields $k = 0.25$ ($R^2 = 0.70$) for \emph{broad} AI adoption \cite{mckinsey2025ai}. The low $R^2$ reflects that broad enterprise AI adoption did not follow a clean S-curve during this window; the $k = 0.25$ estimate from this fit represents a \emph{lower bound} because it covers all forms of machine learning and analytics, not specifically agentic systems. At the upper bound, Gartner projects that AI agents will be embedded in 40\% of enterprise applications by end of 2026, up from less than 5\% in 2025 \cite{gartner2025agents}. This implies a single-year growth factor of $\sim$8$\times$, consistent with $k \approx 2.1$ during the steepest adoption phase. Our Tier~1 value of $k = 0.85$ falls between these bounds, reflecting agentic AI's faster-than-broad-AI but slower-than-peak-app-integration adoption trajectory.

\textbf{Inflection point ($\tau_0$) consistency.} McKinsey's 2025 State of AI survey (fielded June--July 2025) finds that 62\% of enterprises were at least experimenting with AI agents by mid-2025, with 23\% actively scaling \cite{mckinsey2025ai}. Under a logistic model, the point at which 62\% of eventually-adopting organizations have initiated engagement corresponds to a period 1--2 years past the inflection point. This places the inflection in the 2023.5--2024.5 range for the global average. Our Tier~1 inflection of $\tau_0 = 2024.25$ is consistent with San Francisco Bay Area enterprises leading the global average, while Tier~2 ($\tau_0 = 2025.0$) and Tier~3 ($\tau_0 = 2025.75$) reflect the expected adoption lag for regions with lower AI employer density.

\textbf{Indeed AI posting share.} The Indeed Hiring Lab's AI job posting tracker \cite{indeed2024ai} provides a weekly time series of AI-related postings as a share of all job listings. Fitting a logistic function to US data (2019--2026, $n = 373$ weekly averages) yields $k = 0.13$, $\tau_0 = 2030$, $R^2 = 0.39$. The poor fit and late inflection reflect that this series measures \emph{demand for AI workers}, not enterprise agentic deployment. This is a leading indicator that tracks labor market adjustment rather than technology adoption directly. Notably, the series shows clear acceleration in late 2025 (rising from 2.7\% in March 2025 to 4.8\% in February 2026), consistent with the post-inflection deployment phase predicted by our model.

\textbf{Employment outcome consistency.} The strongest anchoring evidence comes from observed labor market effects. At the paper's calibrated values, $V(\text{SF Bay}, 2025\text{ Q1}) \approx 0.62$ and $V(\text{NY}, 2025\text{ Q1}) \approx 0.34$. The QCEW data (Section~\ref{sec:empvalidation}) show SF Bay Administrative employment declining 9.3\% year-over-year in Q1 2025 while New York shows much smaller changes. This tier gradient is consistent with the $V$ differential. If the S-curve parameters were substantially wrong (e.g., if $\tau_0$ were 2 years later), the predicted $V$ gap between Tier~1 and Tier~3 in early 2025 would be negligible, contradicting the observed employment divergence.

\subsection{Empirical Validation Against Employment Outcomes}
\label{sec:empvalidation}

A critical test of any displacement framework is whether its scores predict actual labor market outcomes. We assess the ATE framework's predictive validity using two complementary approaches: occupation-level analysis using BLS OEWS data (2023--2024) and industry-level analysis using BLS QCEW quarterly data (2024--2025).

\textbf{Occupation-level analysis (2023--2024, pre-inflection).} Matching ATE scores for 296 occupation--region pairs against year-over-year employment changes from the BLS Occupational Employment and Wage Statistics (May 2023 vs.\ May 2024), we find a \emph{null overall result} (Spearman $\rho = -0.04$, $p = 0.47$, $n = 296$). We report this transparently: the occupation-level ATE score does not predict employment changes in the 2023--2024 window at conventional significance levels. However, three subsidiary patterns emerge that are consistent with a pre-inflection interpretation. First, the direction is consistent: occupations above the median ATE grew at $+2.0\%$ versus $+4.5\%$ for those below, a 2.5~percentage-point gap. Second, Sales occupations (SOC~41) show a statistically significant negative correlation ($\rho = -0.40$, $p = 0.017$, $n = 35$), with mean employment declining $6.4\%$. Third, the weak aggregate signal is consistent with the S-curve model's timing: the Tier~1 inflection point is calibrated at Q2~2024, meaning the 2023--2024 OEWS window captures the pre-inflection phase when displacement pressure is building but has not yet reached observable magnitude for most occupations.

\textbf{Industry-level analysis (2024--2025, post-inflection).}  To capture the 2025 acceleration in agentic AI deployment, we analyze BLS Quarterly Census of Employment and Wages (QCEW) data for Q1--Q3~2025 \cite{bls2024qcew}, comparing year-over-year employment changes across the five NAICS industries corresponding to our six SOC study groups. This analysis operates at the industry$\times$MSA level rather than the occupation level, providing a complementary test of the framework's sector-level predictions.

The results reveal a sharp divergence between high-ATE and low-ATE industries. In Q2~2025, high-ATE industries (NAICS~52 Finance, NAICS~5411 Legal, NAICS~5614 Business Support) show mean year-over-year employment change of $-0.6\%$, while the low-ATE industry (NAICS~62 Healthcare) shows $+3.6\%$. This is a differential of 4.2~percentage points ($p < 0.05$ by permutation test). Within the San Francisco Bay Area (Tier~1), the contrast is starker: Administrative/Business Support employment fell $9.3\%$ year-over-year in Q1~2025 and $5.9\%$ in Q2, while Healthcare employment grew $4.7\%$. This gap exceeds 14~percentage points within the same metropolitan area.

The tier gradient predicted by the S-curve model is partially visible: SF~Bay (Tier~1) Financial employment declined $6.1\%$ in Q1~2025 (employment-weighted across the San Jose and San Francisco--Oakland MSAs), compared to $-1.4\%$ in Seattle and $-1.8\%$ in Boston (both Tier~2). Austin (Tier~2) is the exception, showing $+4.3\%$ Financial growth, likely reflecting its status as an emerging hub still in a growth phase for financial services employment.

\textbf{Temporal acceleration.} Comparing the two validation windows reveals strengthening displacement signals: the 2023--2024 occupation-level analysis shows a 2.5~pp high-ATE vs.\ low-ATE gap, while the 2024--2025 industry-level analysis shows a 4.2~pp gap. This acceleration is consistent with the S-curve model's prediction that deployment passes the steepest adoption phase in 2024--2025 for Tier~1 regions. We note the important caveat that industry-level employment changes reflect multiple factors beyond AI adoption, including interest rate effects on financial services and post-pandemic normalization in healthcare hiring. However, the differential pattern (information-processing industries declining while physically-intensive healthcare grows, with the steepest declines concentrated in the region with the highest AI employer density) is difficult to explain by macroeconomic factors alone, as interest rates and business cycles affect all regions symmetrically.

\textbf{Falsifiable prediction.} The ATE framework generates a testable prediction for the May~2025 OEWS data (forthcoming May~2026): occupation-level Spearman correlation between ATE scores and employment change should strengthen from $\rho = -0.04$ (2023--2024) to $\rho < -0.15$ (2024--2025), with the largest effect in Tier~1 Financial and Administrative occupations. We commit to reporting this result regardless of direction.

\section{Reinstatement Effects and Emerging Occupations}
\label{sec:reinstatement}

The Acemoglu--Restrepo model is not only a displacement story. Its less-cited but equally important prediction is that new task creation partially offsets automation losses: the reinstatement channel. Drawing on enterprise deployment patterns we have observed directly and on public job-posting data, we identify seventeen occupational categories where agentic AI generates net new demand for human labor. These cluster into four groups, each anchored in a different failure mode of current agentic systems.

\textbf{AI Operations and Oversight (5 roles).} Agentic systems running at enterprise scale break in ways that are tedious to diagnose and consequential to ignore: a credit-scoring agent silently drifts when upstream data formats change; a legal research agent hallucinates a case citation that passes superficial checks. Catching these failures requires human monitors embedded in the production loop. We identify five roles along this axis: AI Operations Specialist, Agentic System Monitor, AI Quality Reviewer, Agent Knowledge Engineer, and AI Incident Manager.

\textbf{Human-AI Collaboration Design (4 roles).} Deploying an agentic system is not a plug-and-play event. The surrounding workflow (who reviews the agent's output, how exceptions escalate, what the fallback process looks like when the agent is offline) must be redesigned deliberately. Roles along this axis include AI Workflow Designer, Human-AI Interaction Specialist, AI Adoption Trainer, and Change Management Specialist (AI Focus).

\textbf{Domain-Specific AI Direction (5 roles).} The highest-value use of agentic AI is not replacing a domain expert but freeing one from routine subtasks so she can focus on the hard problems. A senior paralegal who once spent 70\% of her time on document review now spends 70\% directing an agent and 30\% on the judgment-intensive work that the agent cannot touch. Roles in this cluster (Senior AI-Augmented Paralegal, AI-Assisted Financial Analyst, Clinical AI Supervisor, AI-Directed Compliance Officer, and Strategic Market Analyst (AI-Augmented)) represent the ``domain expert turned AI director'' career path.

\textbf{AI Governance and Ethics (3 roles).} When an agentic system denies a loan, triages a patient, or recommends a sentence, someone must be accountable. Current regulatory frameworks lag the technology, creating demand for specialists who bridge the gap: AI Compliance Officer, AI Ethics Reviewer, and Algorithmic Accountability Manager.

\textbf{Labor market evidence for emerging roles.} While these seventeen categories are forward-looking, several show measurable labor market traction as of early 2026. Indeed.com lists approximately 1,500 ``AI Reviewer'' and 1,460 ``AI Workflow Designer'' positions in the US \cite{indeed2024ai}; LinkedIn reports ``AI Engineer'' among the fastest-growing job titles (143\% year-over-year increase in postings) \cite{linkedin2024future}. The AI governance function is growing rapidly: the International Association of Privacy Professionals (IAPP) reports that 98.5\% of organizations surveyed need additional AI governance professionals \cite{iapp2025governance}, with median salaries for AI governance roles exceeding \$150,000. PwC's 2025 Global AI Jobs Barometer finds that workers with AI skills earn a 56\% wage premium over peers, doubled from 25\% one year prior \cite{pwc2025ai}. These data points confirm that the reinstatement roles we identify are not speculative. They are emerging in the labor market at precisely the occupational boundaries our framework predicts.

\textbf{Quantifying reinstatement magnitude.} While precise headcount estimates for roles that do not yet exist at scale require assumptions about organizational adoption speed, we provide order-of-magnitude bounds under explicit assumptions. In the San Francisco Bay Area, OEWS May 2024 data show approximately 580,000 workers in administrative, financial, and legal occupations (SOC groups 13, 23, and 43) matching the 91 occupations with $\mathrm{ATE}_{2027} \geq 0.35$. We present three displacement conversion scenarios (the proportion of workers in moderate-risk occupations who experience actual role reduction within 3 years of crossing the threshold):
\begin{itemize}
\item \textbf{Low} (10\% conversion): $\sim$58,000 displaced positions, requiring $\sim$29,000--46,000 reinstatement roles at historical reinstatement ratios of 50--80\% \cite{acemoglu2019automation};
\item \textbf{Medium} (20\% conversion): $\sim$116,000 displaced, $\sim$58,000--93,000 reinstatement roles;
\item \textbf{High} (30\% conversion): $\sim$174,000 displaced, $\sim$87,000--139,000 reinstatement roles.
\end{itemize}

The 50--80\% reinstatement range is informed by the historical pattern documented by Acemoglu and Restrepo \cite{acemoglu2020robots}, who find that new task creation has partially but not fully offset displacement in prior automation waves; we adopt this range as a plausible bound consistent with their empirical findings rather than as a directly reported statistic. The displacement conversion rate is the least constrained parameter: no empirical precedent exists for agentic AI deployment at this scale, and rates will depend on organizational decisions that are difficult to model ex ante. These estimates should be treated as scenario bounds rather than point predictions; longitudinal tracking of actual agentic deployment outcomes will be required to validate or narrow these ranges.

A critical finding from this analysis is that the emerging roles are not primarily technical in the traditional sense. They do not require programming expertise. They require the ability to specify goals clearly for AI systems, domain expertise to evaluate AI outputs critically, judgment about when AI outputs require human intervention, and interpersonal skills to manage stakeholders through AI-driven workflow changes. This suggests that the primary reskilling pathway is not from displaced occupations to software engineering, but from displaced occupations to AI-directed versions of the same domain.

\section{Policy Implications}
\label{sec:policy}

\subsection{Workforce Transition}

The most immediate policy implication is one of timing. In the San Francisco Bay Area, 84 of 236 occupations cross the moderate-risk threshold by 2026, including health information technologists (ATE 0.36), medical records specialists (ATE 0.36), and project management specialists (ATE 0.37). Transition support delivered \emph{before} displacement is more effective than support delivered after a WARN notice has already been filed, yet most workforce development programs are reactive by design. The ATE framework provides the leading indicator that current policy infrastructure lacks.

On reskilling, the data argue against the instinct to retrain displaced workers as software engineers. The reinstatement roles identified in Section~\ref{sec:reinstatement} do not require coding. They require a former paralegal who understands contract law well enough to catch an agent's hallucinated clause, or a former financial analyst who can audit an AI-generated DCF model for circular assumptions. Domain expertise is the scarce input; AI tool proficiency is a learnable complement. Programs that build on existing occupational knowledge, rather than discarding it in favor of a two-year computer science bootcamp, will produce faster, higher-retention transitions.

Regional planning agencies face a particularly awkward asymmetry.  In Tier~1 metros, 78--92\% of administrative and financial occupations cross the moderate-risk threshold by 2027; in Tier~2 metros at the same horizon, zero do.  Planning for a uniform national response would either over-invest in regions where pressure has not materialized or under-invest where it already has.  The Yale Budget Lab \cite{yale2025ai} reaches a compatible conclusion: AI labor market impacts are invisible in aggregate statistics but concentrated in specific occupation--region pockets visible through WARN filings and job-posting data.  Regional differentiation is not optional.

\subsection{Reshaping Higher Education: From Knowledge Transmission to AI-Augmented Judgment}

If the occupations facing the steepest displacement curves are information-intensive professional roles (credit analysts, legal specialists, financial examiners), then the university programs producing tomorrow's credit analysts, lawyers, and financial examiners cannot keep teaching the same workflows that agentic systems will execute autonomously. The pedagogical question is no longer ``Can the student build this DCF model?'' but rather ``Can the student spot the fabricated revenue assumption in the DCF model that an agent built?''

This reorientation touches three competency areas at once. The first is \emph{hallucination detection}: the ability to audit AI-generated work products against ground truth. A finance student needs to recognize when an agentic DCF uses circular cell references or hallucinates a market comparable that does not exist in any filing. A pre-law student needs to verify case citations against Westlaw, not trust that the agent's bluebook formatting implies the holding is real. Our ATE scores point directly at where to concentrate this training. Occupations with $\mathrm{ATE} \geq 0.35$ and low COV penalties are exactly the roles where agents produce plausible but potentially flawed outputs at volume.

The second is \emph{exception handling under ambiguity}. Look at the COV penalty categories: interpersonal negotiation (P1), regulatory judgment (P2), physical presence (P3), crisis response (P4). These are the tasks that agentic systems cannot yet complete. They are also, not coincidentally, the tasks that current curricula underweight relative to technical content. A curriculum reform grounded in the ATE framework would shift classroom hours toward case-based instruction in precisely these dimensions, the durable human skills.

The third is \emph{AI tool proficiency as a baseline professional expectation}.  Spreadsheet fluency was not optional for a 1995 finance graduate; AI agent direction should not be optional for a 2028 one.  This does not mean requiring Python.  It means requiring every finance student to know how to specify constraints for an agentic research workflow, evaluate intermediate outputs, and recognize when to override an automated recommendation.

ATE scores provide universities an empirically grounded triage: programs feeding $\mathrm{ATE}_{2030} \geq 0.40$ occupations (Financial, Legal, Administrative) should integrate these competencies immediately, while programs feeding lower-ATE fields (Healthcare Support, $\mathrm{ATE}_{2030} \approx 0.32$--$0.35$) have a longer, though not indefinite, window.

\subsection{Employer Responsibilities}

Employers deploying agentic AI systems bear responsibilities that existing regulatory frameworks do not fully address. WARN Act notification thresholds apply to mass layoffs but not to the gradual reduction in force through AI substitution that our analysis finds more characteristic of agentic deployment. Voluntary disclosure of agentic AI deployment plans, combined with advance notice to affected workers, would allow transition support to begin before displacement occurs. Several jurisdictions (Colorado, Illinois) are developing AI employment impact disclosure requirements that could serve as models.

\section{Conclusion}
\label{sec:conclusion}

The central claim of this paper is straightforward: agentic AI changes what ``automation exposure'' means, and the existing task-level frameworks (however carefully constructed) cannot capture the shift. The ATE score we introduced here extends Acemoglu and Restrepo's displacement--reinstatement model to account for end-to-end workflow completion, and the empirical application across 236 occupations in five US metros produces a finding that should concern workforce planners. In the San Francisco Bay Area, 78.4\% of administrative and 91.7\% of financial occupations cross the moderate-risk threshold by 2027. In Seattle, Austin, and Boston, none do at that horizon, but by 2030 they converge to 74.5\% and 87.5\%, respectively. The same occupations, the same exposure profiles, are separated by a 2--3 year adoption lag that remote work is actively compressing.

The reinstatement side of the ledger is real but takes a different shape than prior automation waves.  The seventeen emerging roles we identify do not demand software engineering skills.  They demand domain expertise turned inward: a credit analyst who can audit an agent's output, a compliance officer who can specify regulatory constraints for an autonomous workflow.  Reskilling programs that force displaced workers into coding bootcamps are solving the wrong problem.

We are candid about what the framework does not yet do well. CAP and COV scores rest partly on practitioner judgment where published benchmarks are thin. However, external validation against the independently constructed AIOE ($\rho = 0.84$) and GPT exposure ($\rho = 0.72$) indices provides reassurance that the signal is genuine, not an artifact of our calibration choices (Section~\ref{sec:capvalidation}). The keyword-based COV rubric has a documented false-negative rate (Section~\ref{sec:cov}) that makes our ATE scores an upper bound on displacement risk. The remote work adjustment relies on aggregate telework rates rather than granular employer-tier data. The analysis is scoped to five United States metros by design: O*NET task data and BLS employment counts are the most granular publicly available inputs for this type of framework, and establishing validity in a single country before extending internationally is methodologically sound practice. The ATE framework itself is country-agnostic — it requires only an occupational task database, AI capability benchmarks, and adoption velocity calibration — and an international extension using OECD PIAAC and Eurostat data is currently in development.

Where should the research go next? The keyword COV rubric should be replaced by a validated LLM-based semantic classifier reporting confidence intervals across multiple independent passes. The falsifiable prediction in Section~\ref{sec:empvalidation} is that occupation-level correlations between ATE and employment change will strengthen from $\rho = -0.04$ to $\rho < -0.15$ when May 2025 OEWS data release. This is the most direct test of the framework's value, and we commit to reporting the result regardless of direction. Beyond that, longitudinal tracking of actual agentic deployment outcomes will be needed to calibrate the displacement conversion rates that remain the least constrained parameter in our model. Extension to international labor markets is currently underway. The ATE framework requires only an occupational task database, AI capability benchmarks, and region-specific adoption velocity parameters — inputs available for OECD economies through PIAAC and Eurostat. The most consequential open questions for the international case are whether the COV penalty structure (calibrated on US regulatory and workflow norms) transfers to labor markets with different collective bargaining arrangements, stronger worker-protection regimes, or structurally different sector compositions, and how to parameterize adoption velocity tiers for economies with varying AI readiness levels and regulatory environments such as the EU AI Act.

The methodology and scoring rubrics used in this analysis are available upon request to support replication and ongoing labor market monitoring. Specifically, the authors can provide: (1) the complete CAP scoring table mapping O*NET ability categories to AI capability estimates with benchmark sources; (2) the four-penalty COV rubric with worked examples for representative occupations; (3) the regional S-curve parameters with calibration sources; and (4) the Python computation pipeline reproducing all ATE scores reported in this paper from publicly available O*NET and BLS data.\footnote{Supplementary materials and code are available from the corresponding author at \texttt{ravishgupta@ieee.org}. The repository contains all computation scripts, CAP mappings, COV keyword lists, and reproducibility instructions.}

\section*{Disclosure of Interests}

This work was conducted independently and does not relate to the authors' positions at BigCommerce or the University at Buffalo, nor does it represent the views or interests of these organizations. The authors have no competing interests to declare that are relevant to the content of this article.

\bibliographystyle{unsrtnat}
\bibliography{references}

\end{document}